\documentclass{article}
\usepackage[utf8]{inputenc}
\usepackage{geometry}
\usepackage{graphicx}
\usepackage{hyperref}
\usepackage{siunitx}
\usepackage{authblk}
\usepackage{amsmath}
\usepackage{float}
\usepackage{color}
\usepackage{bm}

\usepackage{xcolor}

\usepackage[sorting=none]{biblatex}
\begin{document}

%%%%%%%%%%%%%%%%%%%%%%
%%% TITLE AND AUTHORS
%%%%%%%%%%%%%%%%%%%%%%

% Title
\title{Perspectives: Comparison of Deep Learning Segmentation Models on Biophysical and Biomedical Data}

% Authors
\author[1,2]{J. Shepard Bryan IV}
\author[1,2]{Pedro Pessoa}
\author[2, *]{Meysam Tavakoli}
\author[1,3,*]{Steve Press{\'e}}

% Affiliations
\affil[1]{Department of Physics, Arizona State University}
\affil[1]{Center for Biological Physics, Arizona State University}
\affil[3]{Department of Radiation Oncology, Emory School of Medicine}
\affil[4]{School of Molecular Sciences, Arizona State University}
\affil[*]{meysam.tavakoli@emory.edu} 
\affil[*]{spresse@asu.edu} 

% Make title
\maketitle

%%%%%%%%%%%%%%%%%%%%%%
%%% ABSTRACT
%%%%%%%%%%%%%%%%%%%%%%
\begin{abstract}
    Deep learning based approaches are now widely used across biophysics to help automate a variety of tasks including image segmentation, feature selection, and deconvolution. However, the presence of multiple competing deep learning architectures, each with its own unique advantages and disadvantages, makes it challenging to select an architecture best suited for a specific application. As such, we present a comprehensive comparison of common models. Here, we focus on the task of segmentation assuming the typically small training dataset sizes available from biophysics experiments and compare the following four commonly used architectures: convolutional neural networks, U-Nets, vision transformers, and vision state space models. In doing so, we establish criteria for determining optimal conditions under which each model excels, thereby offering practical guidelines for researchers and practitioners in the field.
\end{abstract}

%%%%%%%%%%%%%%%%%%%%%%
%%% INTRODUCTION
%%%%%%%%%%%%%%%%%%%%%%
\section{Introduction}
%%%%%%%%%%%%%%%%%%%%%%
%%% INTRODUCTION
%%%%%%%%%%%%%%%%%%%%%%

Deep learning is increasingly used in biophysics for a variety of tasks~\cite{alquraishi2021differentiable} including time series analysis~\cite{granik2019single,mitchell2024topological}, image reconstruction~\cite{duan2022deep,jin2020deep}, predicting protein structure and dynamics~\cite{pakhrin2021deep,mardt2018vampnets}, drug discovery~\cite{chen2018rise,gawehn2016deep,rifaioglu2019recent}, and segmentation~\cite{falk2019u,morelli2021automating,lavitt2021deep}. Here we focus on time-independent segmentation. Deep learning based segmentation has proven to be a powerful tool, enabling significant advancements in data analysis and automation~\cite{bishop2023deep,goodfellow2016deep}. For instance, it is essential for automatic cell boundary identification used in cell counting~\cite{falk2019u,morelli2021automating,lavitt2021deep}. 

When we refer to a deep learning model, we mean the object or function which takes in data and outputs a prediction. When constructing a deep learning model there are a variety of architectures to choose from, where an architecture is the underlying mathematical structure of a model. Despite the availability of numerous deep learning architectures, each possessing unique strengths and weaknesses, the selection of an appropriate segmentation architecture for specific applications remains challenging. This is especially true when reliable hand-annotated training data is limited, as is often the case in biophysical experiments where a few hundred images exist in contrast to the millions used for more general use cases~\cite{han2022survey}. A systematic and robust evaluation of these models on biophysics data is therefore  essential in providing clear guidelines on their optimal usage.

Model comparison is a core part of deep learning~\cite{goodfellow2016deep,bishop2023deep}. Indeed, various fields have compared models within the specific constraints of each domain where different use cases and architectures have been investigated for electron microscopy~\cite{aswath2023segmentation,horwath2020understanding}, cell segmentation~\cite{pham2018cell,lavitt2021deep}, and fundus imaging~\cite{soomro2019deep, chen2021retinal}. However, there are few systematic reviews covering segmentation in biophysics, especially for small datasets.

Here, we present a comparison and guide for deep learning segmentation networks, focusing on applications to biophysics and medical imaging under small data conditions. Our work addresses this need by comparing the performance of these models on three distinct datasets: phase-contrast imaging of \textit{Bdellovibrio bacteriovorus}~\cite{xu2023two,kaggle_bdello}, fluorescence microscopy imaging of mice neuron cells~\cite{morelli2021automating,hitrec2019neural,kaggle_neuron}, and fundus images of retinas~\cite{jin2022fives}. We compare four prominent architectures: Convolutional Neural Networks (CNNs)~\cite{goodfellow2016deep,bishop2023deep}, U-Nets~\cite{ronneberger2015u}, Vision Transformers (ViTs)~\cite{dosovitskiy2020image, vaswani2017attention, khan2022transformers}, and Vision State Space Models (VSSMs)~\cite{gu2023mamba,xu2024survey}. 

In our comparison, we test each model, quantifying results with standard metrics like model accuracy and specificity but also highlight practical considerations such as the number of parameters and training time. While each architecture has a wide variety of hyperparameters which could be considered during comparison, we will restrict our focus to comparing the base architectures and comparing the number of layers in the networks, relegating comments on additional hyperparameter comparisons for the discussion.

This work is intended to be read side-by-side with our codebase, which can be found on GitHub~\cite{Shepard_BioModelComparison_2024}. In the codebase~\cite{Shepard_BioModelComparison_2024}, we create detailed tutorials explaining how to implement and run deep learning segmentation models with PyTorch. Below we provide an overview of the datasets, architectures, and comparison results. Our results show that no single model outperforms all across scenarios. We conclude with a discussion on criteria for choosing the best architecture for projects in biophysics.

%%%%%%%%%%%%%%%%%%%%%%
%%% DATASETS
%%%%%%%%%%%%%%%%%%%%%%
\section{Datasets}
%%%%%%%%%%%%%%%%%%%%%%
%%% DATASETS
%%%%%%%%%%%%%%%%%%%%%%

Here we introduce each dataset including their relevance to biophysics and unique anticipated obstacles to segmentation. Each dataset is collected under a different type of experiment (implying unique detector properties), and objects of interest. All datasets were labeled by humans. Additionally, each dataset is relatively small, containing at most a few hundred images, in contrast to typical deep learning datasets outside biophysics which can have millions of labeled images~\cite{han2022survey}.

% Data Example Figure
\begin{figure}[H]
    \centering
    \includegraphics[width=\linewidth]{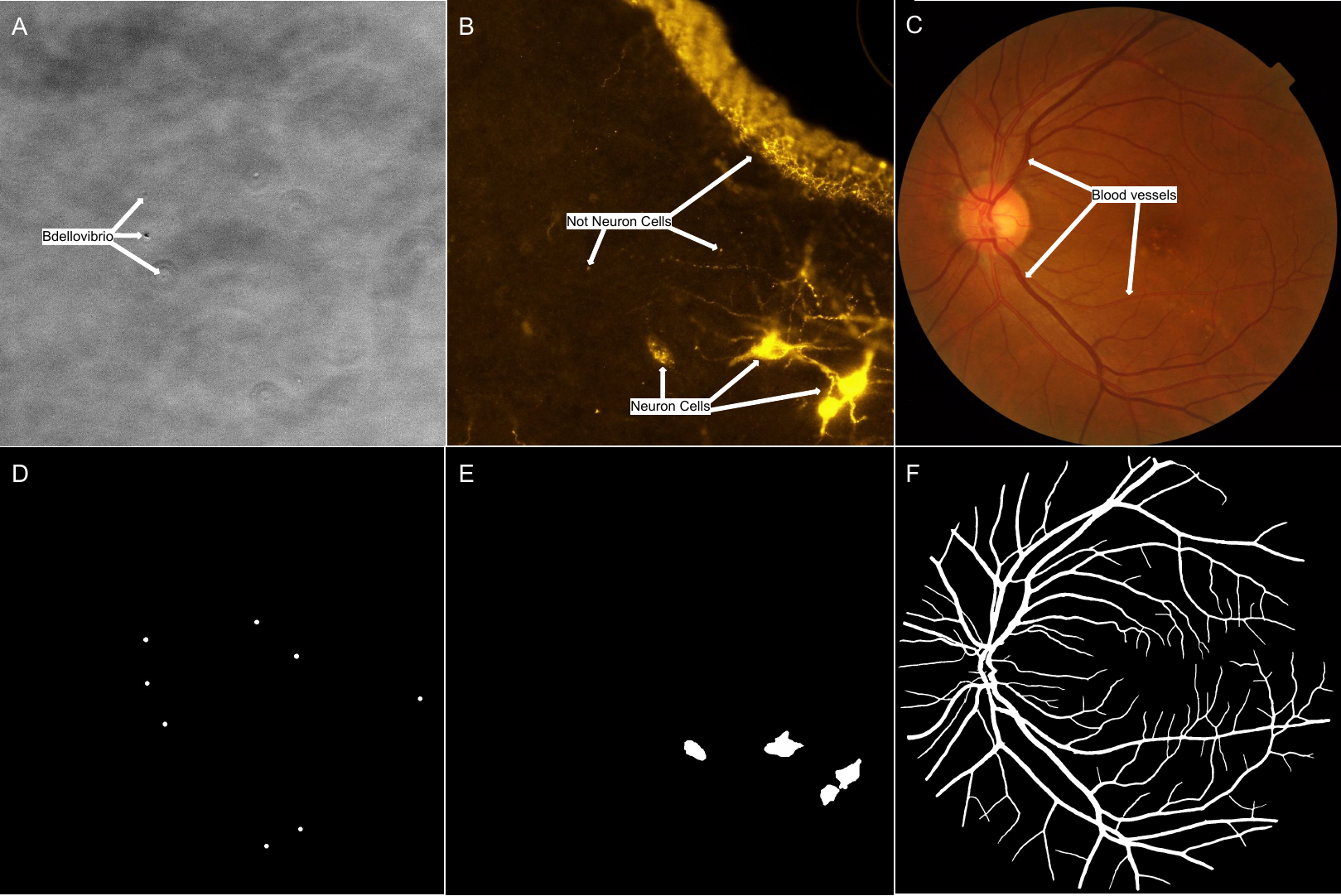}
    % Source image at https://docs.google.com/drawings/d/1xS4Z6CFl7lVoedmitfr01KVGSk-whV2plv0Jckjsg2o/edit?usp=sharing
    \caption{
    \textbf{Example images from each dataset.} Panel A shows an example image from phase-contrast on the \textit{Bdellovibrio} dataset~\cite{kaggle_bdello}, with arrows pointing to Airy rings from a few \textit{Bdellovibrio}. Panel B shows an example image from the fluorescence microscopy mouse neurons dataset~\cite{kaggle_neuron} with arrows indicating the target neuron cells as well as bright objects in the background that the model must learn to ignore. Panel C shows an example retina fundus image~\cite{jin2022fives} with arrows pointing to blood vessels. Panels D, E, and F show the target mask for phase-contrast \textit{Bdellovibrio}, fluorescence microscopy neurons, and retina vessel datasets respectively.
    }
    \label{fig:data_example_images}
\end{figure}

%%% Phase-contrast Bdello
\subsection{Phase-contrast \textit{Bdellovibrio}}
%%% Phase-contrast Bdello

Phase-contrast imaging~\cite{shaked2012biomedical,mehta2022quantitative} is a microscopy technique used to visualize transparent samples by exploiting differences in their refractive indices. Samples are placed under a microscope and illuminated, often with a laser. Differences in phase between the light passing through the sample and a reference beam, caused by variations in the refractive index within the sample, generate contrasting dark and bright regions in the image. This contrast enhancement allows us to visualize otherwise invisible structures within the sample. Phase-contrast imaging is widely used in biophysics~\cite{cheong2015rapid,osvath2018label,xu2023two}, and thus we include it in our comparison.

Our phase-contrast dataset on the bacterium \textit{Bdellovibrio} originates from recent research investigating the hunting strategies of predatory bacteria~\cite{kaggle_bdello}. It contains over 1000 1024$times$1024 images. Under phase-contrast, \textit{Bdellovibrio} are visible due to ring-like Airy discs, with the bacteria positioned at the center of the disc~\cite{xu2023two}. While the Airy disc's z-dependence allows for potential 3D tracking, our focus is restricted to 2D segmentation. In the dataset, bacteria were manually localized, and circular masks  generated around these localizations to serve as the target masks for segmentation experiments. An example image from our phase-contrast \textit{Bdellovibrio} dataset can be found in figure~\ref{fig:data_example_images} panel A.

Segmenting the phase-contrast \textit{Bdellovibrio} dataset presents specific challenges due to the sparsity of masks and the nature of the data generation process. The masks around \textit{Bdellovibrio} localizations constitute less than 1\% of the total pixels, resulting in highly sparse data. Furthermore, \textit{Bdellovibrio} may appear as either dark or bright spots in the images, with the most reliable indicator of bacterial location being the surrounding Airy ring whose radius may span several pixels. Consequently, an effective segmentation algorithm must integrate information from a region larger than the bacteria themselves to accurately determine their locations rather than focusing on either  darker or brighter regions.

%%% Fluorescence microscopy neurons
\subsection{Fluorescence microscopy neurons}
%%% Fluorescence microscopy neurons

Fluorescence microscopy~\cite{godin2014super,kubitscheck2017fluorescence,lindsay2010introduction,fazel2024fluorescence} is a  technique often involving fluorescent dyes labeling molecules of interest, in order to render molecules visible under laser illumination. Fluorescence microscopy is widely used across biophysics and thus we include it in our analysis.

For our analysis of fluorescence microscopy data, we use a dataset of 283 1600$\times$1200 pixel images of fluorescently labelled mouse neurons~\cite{morelli2021automating,hitrec2019neural,kaggle_neuron}. This dataset consists of high-resolution images of mouse brain slices acquired through fluorescence microscopy in which the mouse Raphe Pallidus regions were tagged with Cholera Toxin subunit b fluorescent microbeads~\cite{hitrec2019neural}. This dataset provides a new testbed for computer vision techniques using deep learning~\cite{morelli2021automating,hitrec2019neural,kaggle_neuron}. In these images, neurons appear as yellowish spots of variable brightness and saturation against a generally darker background. The ground-truth labels for this dataset were generated using a combination of semi-automatic and manual segmentation. The lab initially performed an automatic procedure involving Gaussian blurring and histogram-based thresholding to identify potential neuronal cells. This was followed by manual review and correction by experts to ensure accurate annotations. For the most challenging images, including those with artifacts, filaments, and crowded areas, manual segmentation was performed to provide reliable masks. An example image of the fluorescence microscopy neuron dataset can be found in figure~\ref{fig:data_example_images} panel B.

Segmentation of the mice neuron cells dataset poses several challenges due to the variability in brightness and contrast impacting the overall appearance of the images. Neurons exhibit different degrees of fluorescence and their shapes vary significantly. Additionally, structures, such as brigh neuronal filaments, and non-marked cells are present in the images. Neurons may further clump together or overlap, complicating segmentation. The number of target cells varies widely from image to image, ranging from no stained cells to several dozens. Successful segmentation algorithms must account for the fluid nature of the cells, variations in brightness and contrast, and the presence of artifacts and overlapping structures.

%%% Retina fundus images
\subsection{Retina fundus images}
%%% Retina fundus images

Many retinal disorders originate from the blockage of retinal capillaries, with conditions like neovascularization often following nonperfusion and the closure of small retinal blood vessels~\cite{campochiaro2015molecular}. The closure of retinal vessels triggers the formation of neovascularization, and different aneurysms \cite{tavakoli2017comparing, almotiri2018retinal, tavakoli2021unsupervised}. For this reason, the development of screening systems may help in the early diagnosis and real-time classification of retinal diseases such as Diabetic Retinopathy \cite{tavakoli2013complementary, tavakoli2021automated, ruamviboonsuk2022real}, Age-Related Macular Degeneration \cite{kanagasingam2014progress, hwang2019artificial}, and other abnormalities \cite{mayya2021automated, cen2021automatic}. 

Automatic segmentation of ophthalmologic and vascular diseases through retinal image analysis has become a standard practice in healthcare~\cite{badar2020application}. Earlier methods relied on manual segmentation, which was monotonous, time-consuming, inconvenient, labor-intensive, observer-dependent, and required skilled expertise. In contrast, computer-aided segmentation of retinal abnormalities is cost-effective, feasible, more objective, and may avoid the use of highly trained clinicians to evaluate the images~\cite{fraz2012blood}. Many of the retinal vessel segmentation approaches are based on convolutional neural network (CNN) architectures~\cite{liskowski2016segmenting, jiang2019automatic, soomro2019strided, jiang2019retinal}. A comprehensive review of existing methods in retinal vessel segmentation and available
public datasets are presented in~\cite{soomro2019deep, chen2021retinal}. Given the significance and widespread availability of vessel segmentation of retina fundus images, we chose to include them in our study. 

Our study uses the fundus image vessel segmentation~\cite{jin2022fives} (FIVES) dataset, which comprises 800 2048$\times$2048 high-resolution color fundus photographs, meticulously annotated for pixelwise vessel segmentation. The images in this dataset capture both normal eyes and three different eye diseases. The pixelwise annotations were performed through a standard crowdsourcing approach involving trained medical staff and were subsequently verified by experienced ophthalmologists~\cite{jin2022fives}. Each image in the dataset was evaluated for quality using an automatic algorithm, followed by manual corrections from retinal specialists to ensure the accuracy and usability of the data. An example image from the FIVES dataset can be found in figure~\ref{fig:data_example_images} panel C.

Segmenting the retinal vessel images presents unique challenges due to the intricate structures of the retinal vessels. The dataset's diversity, including images of normal and diseased eyes, only further complicates the segmentation task. Accurate segmentation algorithms must handle variations in vessel appearance, differences in image quality, and the presence of pathological features.

%%%%%%%%%%%%%%%%%%%%%%
%%% METHODS
%%%%%%%%%%%%%%%%%%%%%%
\section{Methods}
%%%%%%%%%%%%%%%%%%%%%%
%%% METHODS
%%%%%%%%%%%%%%%%%%%%%%

In this section, we outline the methods employed in this study. We begin with a detailed explanation of the architecture of each model under investigation. We then discuss the metrics used for evaluating model performance. Finally, we provide a summary of our training procedures.

%%% MODELS
\subsection{Model Architectures}
%%% MODELS

% Models figure
\begin{figure}[H]
    \centering
    % Editable via https://docs.google.com/drawings/d/1o7rdcvArJ1KjQKimAgmiYvr3lnQqzaRYmmPiQrZbFTU/edit?usp=sharing
    \includegraphics[width=\linewidth]{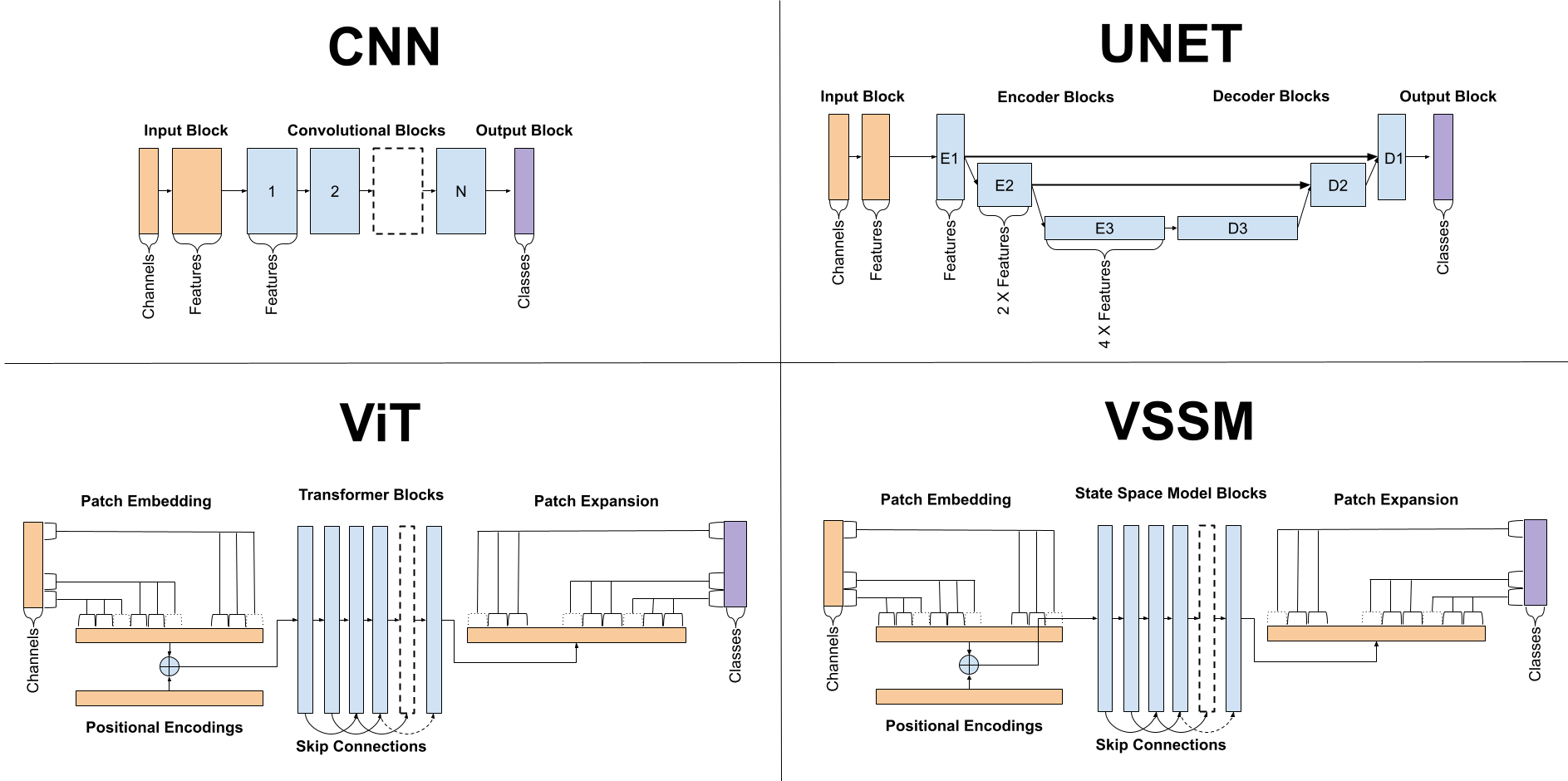}
    \caption{
    \textbf{Diagrams of model architectures.} Here we show rough sketches of the model architectures used in our comparison work. The blocks represent tensors calculated in intermediate steps of the model computation. Blocks with dotted outlines indicate that many blocks are repeated. For example, in the CNN block a dotted box exists between convolutional blocks 2 and N indicating N-2 convolutional blocks in this area. Arrows between blocks correspond to computation.}
    \label{fig:models}
\end{figure}

A segmentation model takes in an image and outputs a new image of the same shape, but with 0s and 1s indicating the class of each pixel, where 1 indicates target and 0 background. This simple framework can be achieved with a variety of different architectures. Here we present an overview of each architecture used in our comparison, highlighting their structure, advantages, and disadvantages. In the companion codebase~\cite{Shepard_BioModelComparison_2024}, we provide examples and tutorials on how to run each model, including the possibility of multi-class segmentation, where there is more than one type of object we wish to segment.

% Convolutional nets
\subsubsection{Convolutional neural network (CNN)}

Convolutional Neural Networks~\cite{goodfellow2016deep,bishop2023deep} (CNNs) operate by applying a kernel (a small filter), over each overlapping patch of an image to extract features. This process, known as convolution, enables CNNs to effectively capture local spatial patterns. The primary advantage of CNNs is their computational efficiency, requiring fewer parameters compared to other architectures discussed herein. However, the local nature of the convolutional operation can be a limitation, as it makes it challenging for CNNs to learn long-range dependencies and interactions within the data.

The top left panel of figure~\ref{fig:models} shows a rough sketch of the block structure of our CNN model. In this diagram we break down the input blocks that preprocess the image, convolutional blocks which process the latent representations of the image, and the output block which merges information into our segmentation output.

% U-nets
\subsubsection{U-net}

U-Nets~\cite{ronneberger2015u} are variants of convolutional networks leveraging skip connections and downsampling to form a ``U''-shaped architecture. This design allows local information to be preserved through skip connections while capturing long-range information via downsampling. During downsampling, the number of feature channels is increased, enabling the network to store more information at each location in the latent space. The advantage of U-Nets over traditional convolutional networks is their ability to access and integrate longer-range information. However, the increasing feature dimensions during downsampling result in a higher number of parameters, making U-Nets generally larger and more memory-intensive.

The top right panel of figure~\ref{fig:models} shows a rough sketch of our U-Net architecture. In this diagram, we show the input block, which preprocesses the image, the encoder block, decoder block, and output block where the segmentation output is finalized. We note that at each block of the encoder the image size is halved (halved along both the width and height so quartered in total) and the feature size is doubled. Also, each encoder is connected to both the subsequent encoder as well as the corresponding decoder via a ``skip'' connection (a connection that skips the intermediate layers).

% Vision transformers
\subsubsection{Vision transformers (ViT)}

Vision Transformers~\cite{dosovitskiy2020image, vaswani2017attention, khan2022transformers,han2022survey} (ViTs) leverage the transformer architecture to process images. In a ViT, the image is divided into a sequence of fixed-size, non-overlapping patches, then linearly embedded into a feature space and augmented with positional encoding vectors. The transformer architecture applies a self-attention mechanism to these patches, allowing each patch to be influenced by information from every other patch. This process enables the model to capture long-range dependencies across the entire image. The advantage of ViTs relies in their ability to distribute long-range information across the network, facilitating the detection of connections between distant regions of the image. However, ViTs typically require large amounts of computational resources and data, as their complexity scales quadratically with the number of patches. This often restricts their application to smaller images (224$\times$224 pixels for the original ViT applications~\cite{dosovitskiy2020image}). Additionally, because ViTs process images in patches, the reconstructed output can exhibit block-like artifacts.

The bottom left panel of figure~\ref{fig:models} shows a rough sketch of the ViT used in our comparison. We note that the input block consists of splitting the image into disjoint patches, then reassembling them into a shape compatible with the transformer architecture. After the patch embedding phase, the latent representation is added to positional encodings, which add back location information otherwise removed during the patchification. This output is then passed to the transformer block in which a series of transformers manipulate the image. Notice the skip connections in the transformer block. From here the output is passed to a patch expansion block which assembles the segmentation output from the latent patch embeddings.

% Vision state-space-model
\subsubsection{Vision state-space-model (VSSM)}

Vision State-Space Models~\cite{gu2023mamba,xu2024survey} (VSSMs) are a relatively new type of network similar to Vision Transformers (ViTs) but otherwise designed to avoid the costly quadratic scaling. At each layer of a VSSM, the image is divided into non-overlapping patches and transformed into a feature space. In this feature space, a series of recurrent neural networks (RNNs) are applied, modifying each patch's features based on information from other sequentially spaced patches. After the recurrent networks are applied, the patches are transformed back into image space and combined with the original image. VSSMs theoretically offer the same advantages as ViTs in terms of capturing long-range dependencies but with computational complexity that scales as $N\log(N)$, where $N$ is the number of patches, which can be significantly less than the quadratic scaling of ViTs though the speed advantage of VSSMs is often hardware-specific. In other words, on a regular CPU, this advantage may be minimal. Similar to transformers, the disadvantage of VSSMs lies in their potential for block-like artifacts in the output image.

The bottom right panel of figure~\ref{fig:models} shows a rough sketch of the VSSM architecture. This architecture is equivalent to the ViT architecture, only with transformers replaced by state space models.

%%% METRICS
\subsection{Metrics}
%%% METRICS

Here, we evaluate the performance of the models using both hard and soft metrics. The hard metrics we will compare include accuracy, specificity, sensitivity, and the ``Area Under the Curve''
(AUC) score.

Accuracy, specificity, and sensitivity are convenient linear combinations of the true positives (when prediction and ground truth are both positive), true negatives (when prediction and ground truth are both negative), false positives (when prediction is positive, but ground truth is negative) and false negatives (when prediction is negative, but ground truth is positive). Accuracy, defined as the ratio of correctly predicted instances to the total instances, provides a general measure of how well the model performs across all classes. It is given by:
\begin{align}
    \text{Accuracy} =& \frac{TP + TN}{TP + TN + FP + FN}
\end{align}
where $TP$ is the number of true positives, $TN$ is the number of true negatives, $FP$ is the number of false positives, and $FN$ is the number of false negatives. Specificity, defined as the ratio of correctly predicted negative instances to the total actual negatives, measures the model's ability to correctly identify negative instances. It is given by:
\begin{align}
    \text{Specificity} =& \frac{TN}{TN + FP}.
\end{align}
Sensitivity, also known as recall, defined as the ratio of correctly predicted positive instances to the total actual positives, indicates the model's ability to correctly identify positive instances. It is given by:
\begin{align}
    \text{Sensitivity} = \frac{TP}{TP + FN}.
\end{align}

The AUC score measures the area under the Receiver Operating Characteristic (ROC) curve~\cite{bishop2023deep}. The ROC curve plots the true positive rate (sensitivity) against the false positive rate (specificity) at various threshold settings, providing insights into the trade-offs between sensitivity and specificity. 

In addition to these metrics, we will also plot the ROC curve to visualize the performance of the models across different threshold settings. Further information on the hard metrics used in this analysis are detailed in the companion codebase~\cite{Shepard_BioModelComparison_2024}.

We will additionally plot both the training and validation loss per epoch. When training deep learning models, the expectation is that the loss on the training set will trend downward, with occasional upward fluctuations, indicating that the model is progressively learning to predict the data better over time. Meanwhile, the validation loss will initially decrease as the model improves its generalization (how well the model performs on similar data outside the training set). However, after reaching a minimum point, the validation loss will begin to increase even as the training loss continues to decrease, signaling that the model is overfitting the training data~\cite{goodfellow2016deep}. By plotting and examining these losses, we ensure that the training and validation losses follow this expected pattern. The goal is to achieve a lower and faster inflection in the validation loss, indicating optimal model generalizability.

The first soft metric we will use to compare our models is their model size. Model size, which refers to the number of parameters, correlates with the memory footprint and storage requirements of the model. Although it's possible to create large models from simple layers (e.g., by using thousands of convolutional layers), certain models, such as ViT and VSSM, inherently have more parameters per layer. Since each model presented in this work will have a reasonably similar number of layers (from 2 to 20, but not 1000) the total number of parameters serves as a good proxy for model complexity.

The second soft metric we will use to compare our models is the training time. Training time is related to the speed at which the model converges during gradient descent, as well as the speed of output for a single image during training. Since we train each model on a different node on our HPC~\cite{HPC:ASU23}, the processing speeds of different jobs may be faster or slower, independent of inherent model speed. For example, a slower model might train on a faster node, making its training appear faster. When looking at the training times, we will therefore note rough orders of magnitude.

%%% TRAINING
\subsection{Training}
%%% TRAINING

We implemented each model from scratch using PyTorch and trained each one on each dataset using five-fold cross-validation~\cite{goodfellow2016deep}. In this setup, the dataset is split into five disjoint subsets. For each fold, three subsets are used for training, one for validation, and one for testing. This process is repeated five times, rotating the validation and testing subsets, then the final hard metrics are averaged over all 5 folds to ensure a robust evaluation of model performance. For training, each image was split into $128 \times 128$ cropped regions.

Each model was trained for 100 epochs with the objective of minimizing the binary cross-entropy loss between the predicted outputs and the ground truth masks. During training, the parameter set minimizing the validation loss was saved. We conducted the training on Arizona State University's Sol high-performance computing cluster~\cite{HPC:ASU23}. Although we allocated a week for the training process, many jobs completed much sooner. Jobs that did not converge within the allotted 100 epochs were rerun with 500 epochs, at which point all jobs converged.

Upon completion of training, we selected the model parameter set that demonstrated the best performance for further analysis and demonstration. This comprehensive approach ensured that our results were reliable and reflective of the models' true performance across different datasets.

More details on the training procedure, as well as a detailed tutorial on how to implement it for other datasets, are provided in the companion codebase~\cite{Shepard_BioModelComparison_2024}.

%%%%%%%%%%%%%%%%%%%%%%
%%% RESULTS
%%%%%%%%%%%%%%%%%%%%%%
\section{Results}
%%%%%%%%%%%%%%%%%%%%%%
%%% RESULTS
%%%%%%%%%%%%%%%%%%%%%%

In this section, we present the results of our model comparison. We evaluate the performance of each model on the three selected datasets using the best parameter values obtained from the training phase. Additionally, in the Supplementary Information (SI), we provide a comparison of each model under different parameter regimes to highlight the impact of parameter variations. The results are presented in two formats: First, we include example outputs on randomly sampled images to qualitatively illustrate the effectiveness of each model. Second, we provide quantitative plots showing the training loss over time and the ROC curves to demonstrate the models' performance and their ability to discriminate between different classes.

% Bdello figures
\begin{figure}[H]
    \centering
    \includegraphics[width=\linewidth]{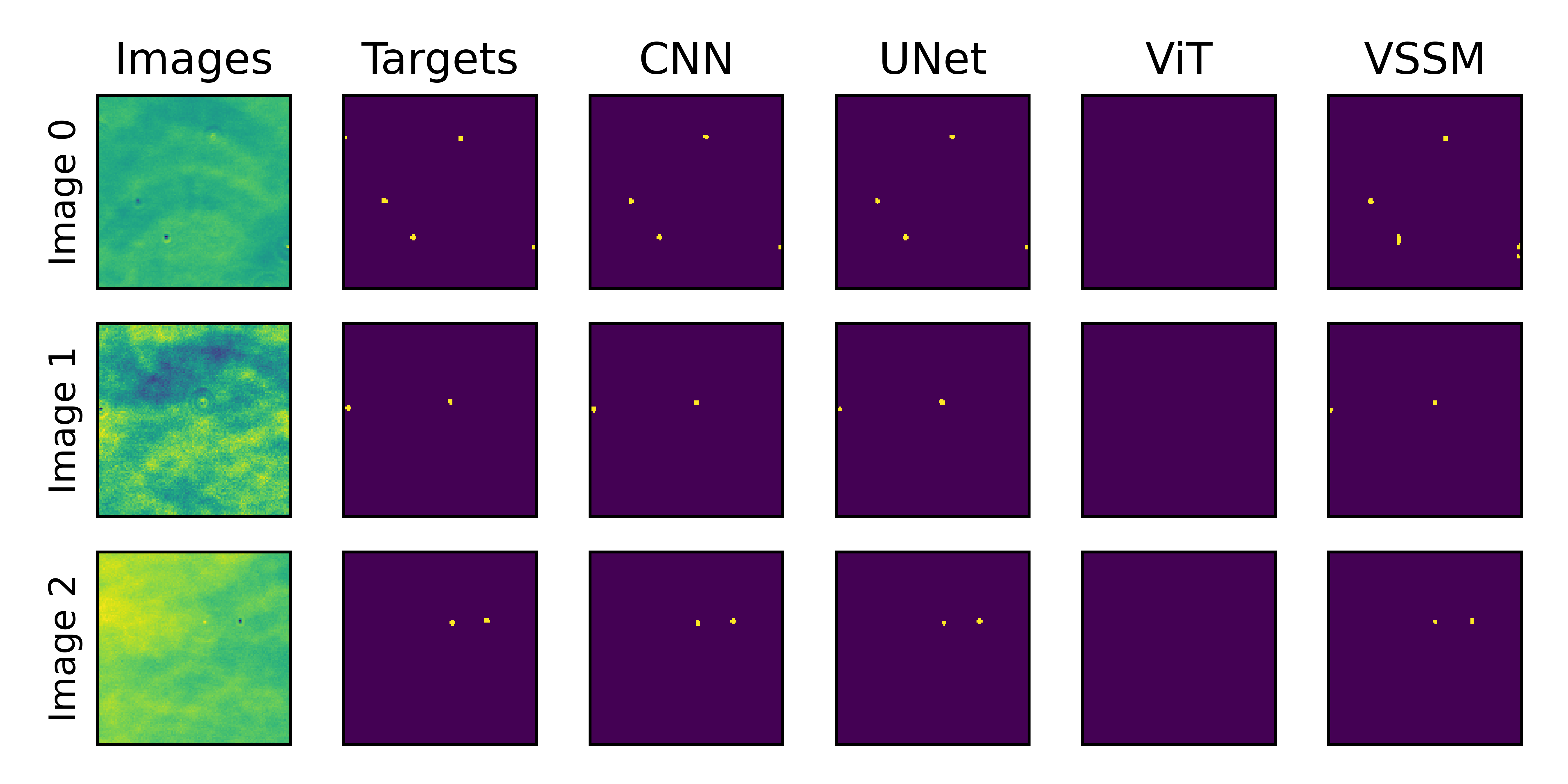}
    \caption{
    \textbf{Example outputs on phase-contrast \textit{Bdellovibrio} data.} The left column shows three input images. The second column shows the target segmentation mask. The remaining columns show the model outputs from each model on each image.}
    \label{fig:bdello_outputs}
\end{figure}
\begin{figure}[H]
    \centering
    \includegraphics[width=\linewidth]{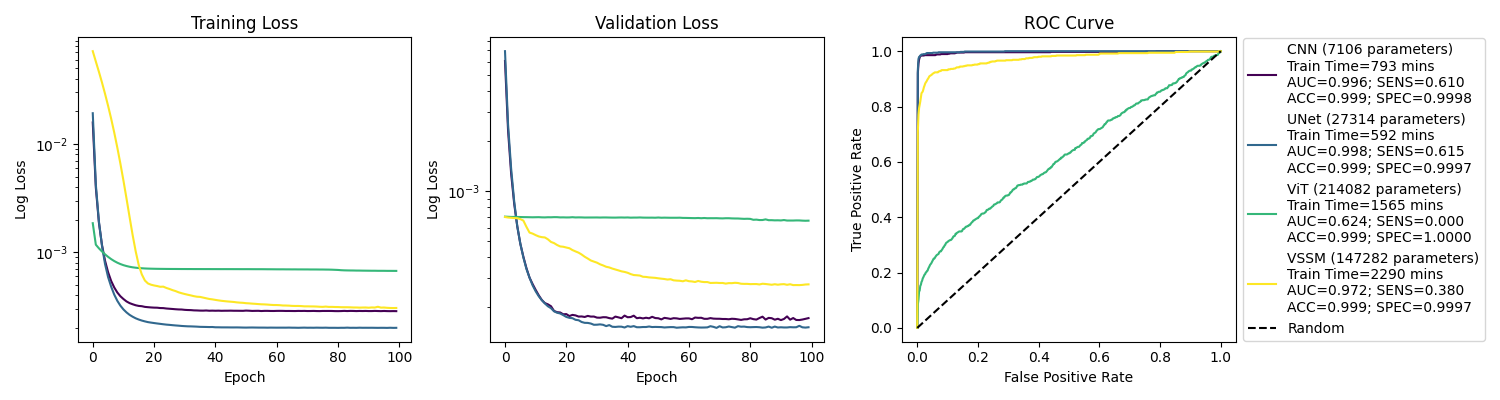}
    \caption{
    \textbf{Training statistics on phase-contrast \textit{Bdellovibrio} data.} The left panel shows training loss per epoch. The middle panel shows validation loss per epoch. The right panel shows the ROC curve. In each panel, we plot the corresponding statistics of each model in different colors. The model corresponding to each color can be found in the legend on the far right. Acronyms are defined in the main text.}
    \label{fig:bdello_stats}
\end{figure}

% Bdello qualitative results
Figure~\ref{fig:bdello_outputs} displays the outputs of our models on the phase-contrast \textit{Bdellovibrio} dataset. Each row corresponds to a different image from the dataset, where the bacteria can appear as either light or dark spots, with the Airy ring around them serving as the most noticeable feature. The two leftmost columns of figure~\ref{fig:bdello_outputs} show the input images and the target masks. The remaining columns show the outputs from our deep learning models. Despite the complexity of the data, the CNN and U-Net models were able to produce qualitatively accurate results. The VSSM was able to identify the locations of the \textit{Bdellovibrio}, but struggled to create correctly shaped target masks, largely due to gridding artifacts from its patchification. The ViT, was not able to identify any bacteria and instead learned to output a target mask in which all pixels were identified as background. This is because the main advantage of the ViT model, an ability to pick up long range correlations, did not outweigh the main disadvantages, gridding artifacts and absence of translational invariance, for detecting the small target bacteria.

% Bdello test metrics
We quantify the hard performance metrics of model performance for phase-contrast \textit{Bdellovibrio} in figure~\ref{fig:bdello_stats} where we present the training loss per epoch, the validation loss per epoch, and the ROC curve for each model. Each model is represented by a distinct color, as indicated in the legend on the far right. The legend also includes the area-under-curve (AUC) score, accuracy (ACC), sensitivity (SENS), and specificity (SPEC). As seen from figure~\ref{fig:bdello_stats}, all models exhibit high accuracy and specificity values, primarily due to the sparsity of bacteria in the dataset. More notably, the CNN and U-Net achieve higher sensitivity and AUC scores, indicating better overall performance on this dataset. While the ViT did not achieve good results, the validation loss shows a sharp downward trend toward the end of the last epoch, indicating that the model may learn to identify \textit{Bdellovibrio} better if trained longer.

% Bdello review
Our results for model performance on the phase-contrast \textit{Bdellovibrio} dataset are in line with expectations as the bacteria are most identifiable by the Airy disc surrounding them, models must integrate information larger than the segmentation area, a situation in which CNNs and U-Nets excel. While the ViT and VSSM in theory allow the model to learn long range interactions, they are limited by their reliance on patches, where each model must learn to identify features of an Airy disc at different orientations and locations. In contrast, CNN and U-Net architectures benefit from their ability to learn a single spatially invariant kernel, enabling more efficient and accurate segmentation.

% Neuron figures
\begin{figure}[H]
    \centering
    \includegraphics[width=\linewidth]{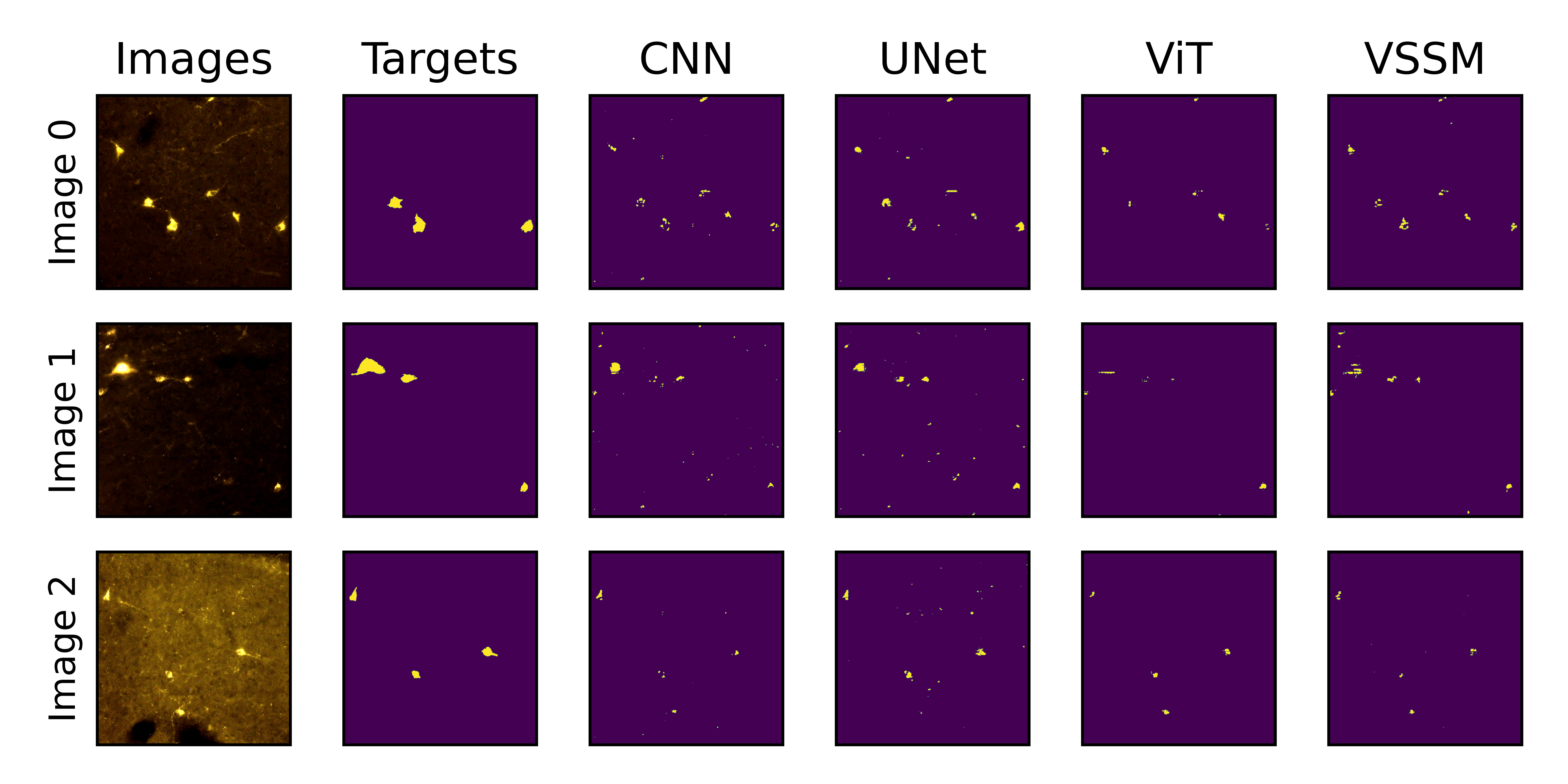}
    \caption{
    \textbf{Example outputs on fluorescence microscopy neuron data.} This image is the same as figure~\ref{fig:bdello_outputs}, but with fluorescent neurons.}
    \label{fig:neurons_outputs}
\end{figure}
\begin{figure}[H]
    \centering
    \includegraphics[width=\linewidth]{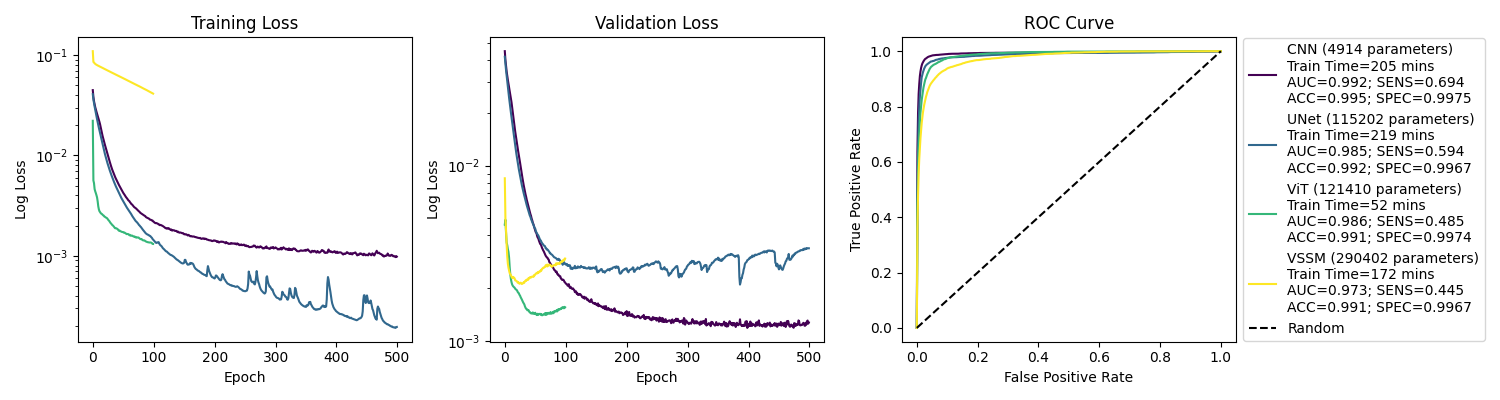}
    \caption{
    \textbf{Training statistics on fluorescence microscopy neuron data.} This image is the same as figure~\ref{fig:bdello_stats}, but with fluorescent neurons.}
    \label{fig:neurons_stats}
\end{figure}

% Neuron qualitative results
Figure~\ref{fig:neurons_outputs} shows the outputs of each model on the fluorescence microscopy mouse neuron dataset. As with the previous dataset, each row in figure~\ref{fig:neurons_outputs} corresponds to a different input image, while each column represents the input images, target masks, and the outputs from the models. This dataset is notably noisier and more challenging to analyze due to the heterogeneity of the target neurons and the frequent presence of artifacts. Visually, it is evident that all models struggled to accurately segment the target regions. The CNN and U-Net exhibited more false positives, whereas the ViT and VSSM produced fragmented and patchy segmentations of the targets.

% Neurons test metrics
Figure~\ref{fig:neurons_stats} presents the hard performance metrics for each model on the neuron dataset. The CNN, U-Net, and ViT performed similarly in terms of AUC score, accuracy, and specificity. However, the training time for the CNN and U-Net was significantly longer. The CNN and U-Net architectures required more epochs to converge (over 200 epochs) compared to the ViT and VSSM, which converged in fewer than 50 epochs.

% Neuron review
The main challenge in analyzing the fluorescence microscopy mouse neuron dataset is the high frequency of false positives. Models must not only identify regions of higher brightness but also filter out bright regions of different morphologies. The larger parameter space of the ViT and VSSM, along with their ability to embed entire patches into a latent space simultaneously, enables them to learn this dual computation more efficiently. While the CNN and U-Net architectures can eventually achieve similar accuracy by identifying kernels over latent embeddings distinguishing different morphologies, this process requires longer training.

% Retina figures
\begin{figure}[H]
    \centering
    \includegraphics[width=\linewidth]{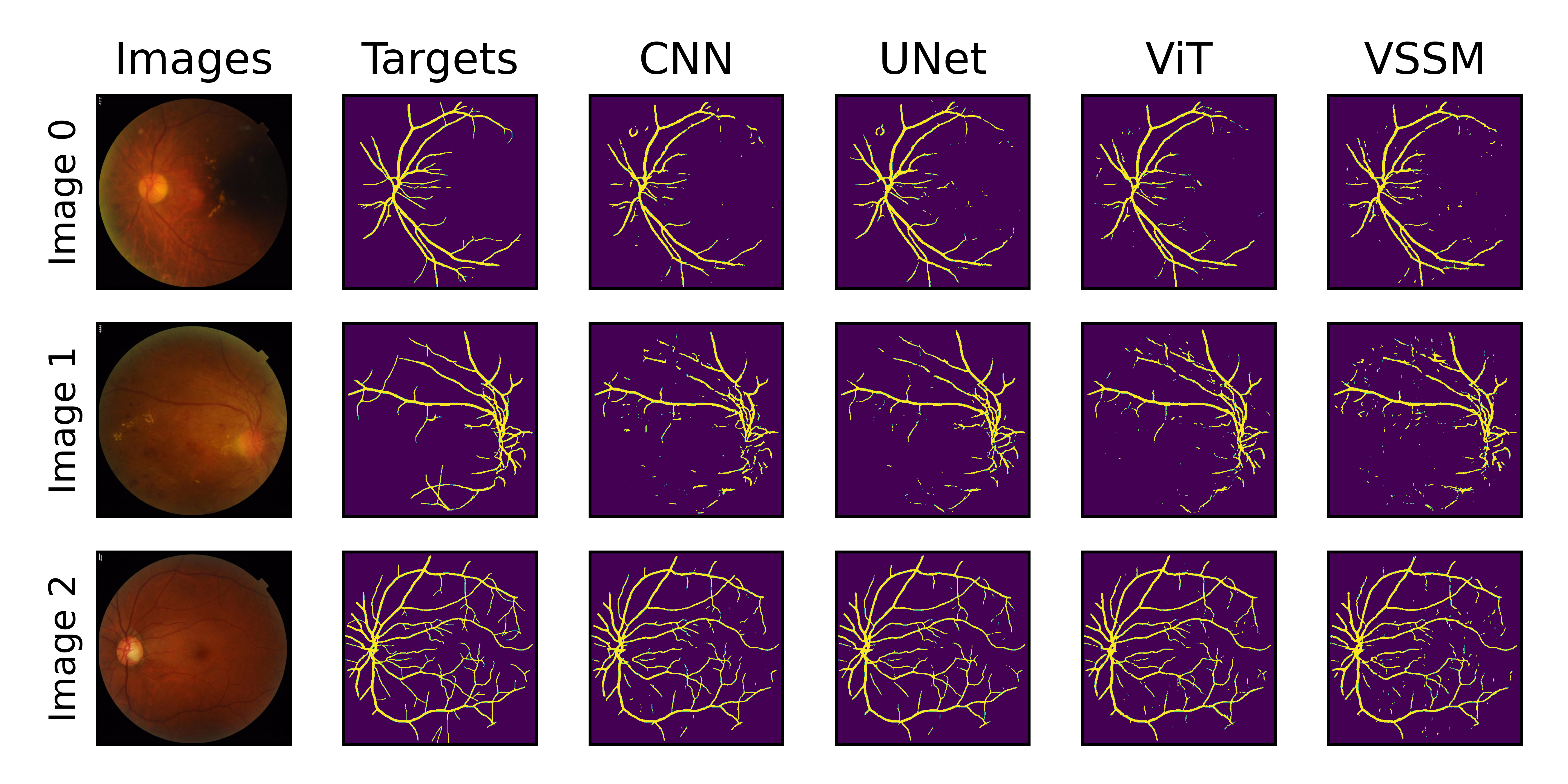}
    \caption{
    \textbf{Example outputs on retina fundus data.} This image is the same as figure~\ref{fig:bdello_outputs}, but with fluorescent neurons.}
    \label{fig:retina_outputs}
\end{figure}
\begin{figure}[H]
    \centering
    \includegraphics[width=\linewidth]{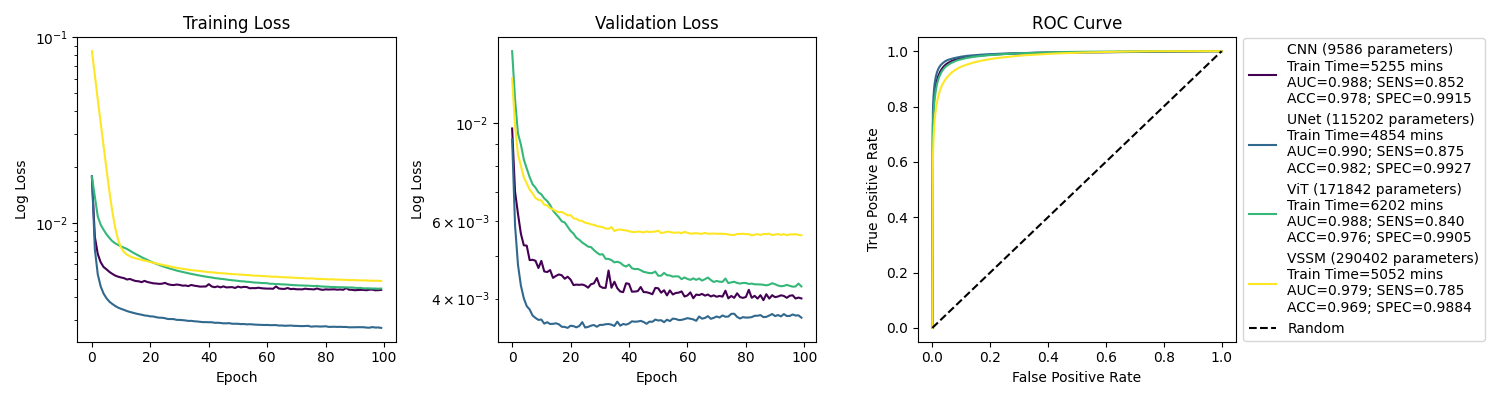}
    \caption{
    \textbf{Training statistics on retina fundus data.} This image is the same as figure~\ref{fig:bdello_stats}, but with retina vessels.}
    \label{fig:retina_stats}
\end{figure}

% Retina qualitative results
Figure~\ref{fig:retina_outputs} displays the outputs of each model on the retina vessel dataset. Similar to the previous datasets, each row in figure~\ref{fig:retina_outputs} corresponds to a different input image, with columns representing the input images, target masks, and the outputs from the models. While all models performed well on this dataset, the U-Net produced qualitatively superior reconstructions, with fewer false positives and clearer, less fragmented segmentations. Given that U-Nets were originally designed for medical segmentation tasks, this result is expected.

% Retina test metrics
Figure~\ref{fig:retina_stats} presents the hard performance metrics for each model on the retina vessel dataset. The data in figure~\ref{fig:retina_stats} show that the U-Net outperforms the other models across all four hard metrics also converging in only 20 epochs compared to over 50 for the other models.

% Retina review
The strong performance of all models on the retina vessel dataset highlights their ability to handle the relatively high-contrast and structured nature of retinal images. The U-Net's architecture, with its skip connections and symmetrical design, excels in preserving fine details and ensuring coherent segmentations, which is why it slightly outperforms the other models in accuracy and AUC. The ViT and VSSM models also perform exceptionally well due to their ability to capture long-range dependencies and integrate global context, crucial in accurately identifying the complex branching patterns of retinal vessels. The CNN, with its straightforward approach, manages to achieve competitive results by effectively learning spatial hierarchies. However, the U-Net's specialized design for medical image segmentation allows it to strike the best balance between local and global information, resulting in fewer false positives and less fragmented segmentations. Thus, while  differences are subtle, the U-Net's architecture provides a slight edge in performance for this particular task.

% Summary table
\begin{center}
\begin{tabular}{|p{3cm}|p{3cm}|p{3cm}|p{3cm}|p{3cm}|}
 %%% Header
 \hline
 & CNN & U-Net & ViT & VSSM\\
 \hline
 %%% Phase-contrast Bdello
 \hline
 \textbf{\textit{Bdellovibrio}} &&&&\\
 AUC Score & 
 .996 & .998 & .624 & .972 \\
 Accuracy & 
 .999 & .999 & .999 & .999 \\
 Sensitivity & 
 .610 & .615 & 0.0 & .380 \\
 Specificity & 
 .9998 & .9997 & 1.0 & .9997 \\
 Parameters & 
 7206 & 27314 & 214082 & 147282 \\
 Training Time & 
 793 mins & 592 mins & 1565 mins & 2290 mins \\
 %%% Fluorescence microscopy neurons
 \hline
 \textbf{Neurons} &&&&\\
 AUC Score & 
 .992 & .985 & .986 & .973 \\
 Accuracy & 
 .995 & .992 & .991 & .991 \\
 Sensitivity & 
 .694 & .594 & .485 & .445 \\
 Specificity & 
 .9975 & .9967 & .9974 & .9967 \\
 Parameters & 
 4914 & 115202 & 121410 & 290402 \\
 Training Time & 
 205 mins & 219 mins & 52 mins & 172 mins \\
 \hline
 %%% Retina vessesls
 \hline
 \textbf{Retinas} &&&&\\
 AUC Score & 
 .988 & .990 & .988 & .979 \\
 Accuracy & 
 .979 & .982 & .976 & .969 \\
 Sensitivity & 
 .852 & .875 & .840 & .785 \\
 Specificity & 
 .9915 & .9927 & .9905 & .9884 \\
 Parameters & 
 9586 & 115202 & 171842 & 290402 \\
 Training Time & 
 5255 mins & 4854 mins & 6202 mins & 5052 mins \\
 \hline
 %%% ADVANTAGES
 \hline
 \textbf{Advantages}
 & % CNN
 Few parameters, fast analysis
 & % U-Net
 Designed to integrate information from multiple length scales
 & % VIT
 Allows long range connections
 & % VSSM
 Allows long range connections, fast on GPUs
 \\ \hline
 %%% DISADVANTAGES
 \hline 
 \textbf{Disadvantages}
 & % CNN
 Limited to local kernels
 & % U-Net
 Additional blocks are memory intensive
 & % VIT
 Gridding artifacts
 & % VSSM
 Gridding artifacts, slow without GPUs
 \\ \hline
 %%% RECCOMENDATIONS
 \hline
 \textbf{Recommendation}
 & % CNN
 Useful for most biophysics datasets
 & % U-Net
 Best when large scale structures are distinguishable at short ranges
 & % VIT
 Best when training data has annotation artifacts
 & % VSSM
 Best when ViT is optimal, but computation is slow
 \\ \hline
\end{tabular}
\label{table:summary}
\end{center}

% Summary
We summarize our results in table~\ref{table:summary}. Looking at the results of our analysis, we see that the CNN is a lightweight and simple algorithm, performing well in all presented scenarios. The U-Net has an advantage when the target is large. The ViT and VSSM are much larger than the CNN and U-Net, but provided faster analysis in the fluorescence microscopy neuron dataset, where the training data had numerous artifacts. The VSSM, did not outperform the other methods for any of our test cases, however it gave similar results to the ViT, and was faster than the ViT when deployed on GPU (but was slower when deployed on CPU).

%%%%%%%%%%%%%%%%%%%%%%
%%% DISCUSSION
%%%%%%%%%%%%%%%%%%%%%%
\section{Discussion}
%%%%%%%%%%%%%%%%%%%%%%
%%% DISCUSSION
%%%%%%%%%%%%%%%%%%%%%%

Deep learning has become a pivotal tool in a variety of disciplines, including biophysics, where systematic comparisons are essential for effective model selection. In this study, we conducted a comprehensive comparison of different deep learning models for biophysics data segmentation.

Our findings indicate that for most small biophysical datasets of around a few hundred images (as opposed to millions of images used for other purposes~\cite{han2022survey}), the Convolutional Neural Network (CNN) and U-Net architectures performed best in simple segmentation tasks. Of the two convolution based methods, the U-net had better qualitative results, with fewer false positives and more connected regions. On the other hand, the Vision Transformer (ViT) and Vision State Space Model (VSSM) had slightly better qualitative results on the more complex, fluorescence microscopy. As ViTs typically perform best in large data situations where the number of data reaches millions of examples~\cite{han2022survey}, our comparison on small datasets did not play to its greatest advantage. 

Overall, our recommendation is that when utilizing deep learning for segmentation, researchers should try to implement simple CNNs first, evaluating performance using both qualitative and quantitative metrics. If initial results do not match the desired quality, then more advanced networks should be employed starting with U-Nets, then moving to ViTs. Beyond these, other exotic architectures could be considered, such as a hybrid between convolutional networks and transformers~\cite{gu2023convformer}, which could blend together the best aspects of each network.

While we focused only on comparing base model architectures, with varying number of layers, there are multiple other hyperparameters impacting training and performance that could be compared. For example, ViTs and VSSMs can have different patch sizes~\cite{dosovitskiy2020image}, and all networks considered can be created with different number of latent features~\cite{goodfellow2016deep}. Given the combinatorically large number of hyperparameter combinations we could compare, it makes sense to limit this initial study to basic architectures. This study may be expanded in the future to consider comparisons of different hyperparameters and training protocols.

Another important consideration is the optimizer used in training. Here we used a standard ADAM optimizer~\cite{kingma2014adam}, but other choices such as stochastic gradient descent~\cite{goodfellow2016deep} could also be benchmarked. 

Furthermore, there are other concepts in deep learning, aside from model architectures and optimizers, that may affect model performance. For example, by pre-training the network as a denoising autoencoder (a model that takes in a noisy image and outputs a cleaned version of the same image), a model may learn to detect features faster and more efficiently than training directly as a segmentation tool~\cite{goodfellow2016deep}. 

Lastly, standard deep learning segmentation models typically output a single prediction of pixel classes, making uncertainty quantification difficult. Bayesian deep learning~\cite{wang2020survey} presents a framework in which outputs are accompanied by uncertainty estimates, which can be especially useful in situations where we wish to quantify our confidence in an output. Bayesian deep learning represents a paradigm of deep learning, not immediately tied to any specific model architecture~\cite{wang2020survey}. As such, a similar comparison of deep learning arcitectures within the Bayesian deep learning framework would is also of future interest. 

Beyond segmentation, there are many other types of deep learning paradigms used in biophysics including time series analysis~\cite{granik2019single,mitchell2024topological}, image reconstruction~\cite{duan2022deep,jin2020deep}, predicting protein structure and dynamics~\cite{pakhrin2021deep,mardt2018vampnets}, and drug discovery~\cite{chen2018rise,gawehn2016deep,rifaioglu2019recent}. As the inputs and target outputs are different for each paradigm, so too will be the expected efficacy of different deep learning architectures. CNNs and U-Nets were originally designed for image processing~\cite{goodfellow2016deep}, thus they performed well in our work. On the other hand ViTs and VSSMs were designed originally for time series data~\cite{goodfellow2016deep}, thus it is possible that they would outperform CNNs and U-Nets for other tasks within biophysics. It would be of interest to repeat our comparison work for other domains of deep learning in biophysics.

%%%%%%%%%%%%%%%%%%%%%%
%%% END MATERIAL 
%%%%%%%%%%%%%%%%%%%%%%

% Code availability
\subsection*{Code availability}
All code used in this work, as well as a script to quickly download data, can be found in a repository on our GitHub~\cite{Shepard_BioModelComparison_2024}.

% Acknowledgements
\subsection*{Acknowledgements}
SP acknowledge support from the NIH (grant no. R01GM134426 and R01GM130745) and support from R35GM148237.

% Competing Interests
\subsection*{Competing Interests}
SP and JSB acknowledge a competing interest with their affiliation with Saguaro Solutions.

% Print bibliography
\printbibliography

%%%%%%%%%%%%%%%%%%%%%%%%%%%%%%%
%% SUPPLEMENTARY INFORMATION %%
%%%%%%%%%%%%%%%%%%%%%%%%%%%%%%%
\clearpage
\section*{Supplementary information}
% Put S in front of refs
\makeatletter
\renewcommand \thesection{S\@arabic\c@section}
\renewcommand\thetable{S\@arabic\c@table}
\renewcommand \thefigure{S\@arabic\c@figure}
\makeatother
% Reset counters
\setcounter{equation}{0}
\setcounter{section}{0}
\setcounter{subsection}{0}
\setcounter{figure}{0}

%% Comparing model parameters
\subsection{Comparing model parameters}
\label{SI:sec:Comparing_model_parameters}

\begin{figure}[H]
    \centering
    \includegraphics[width=\linewidth]{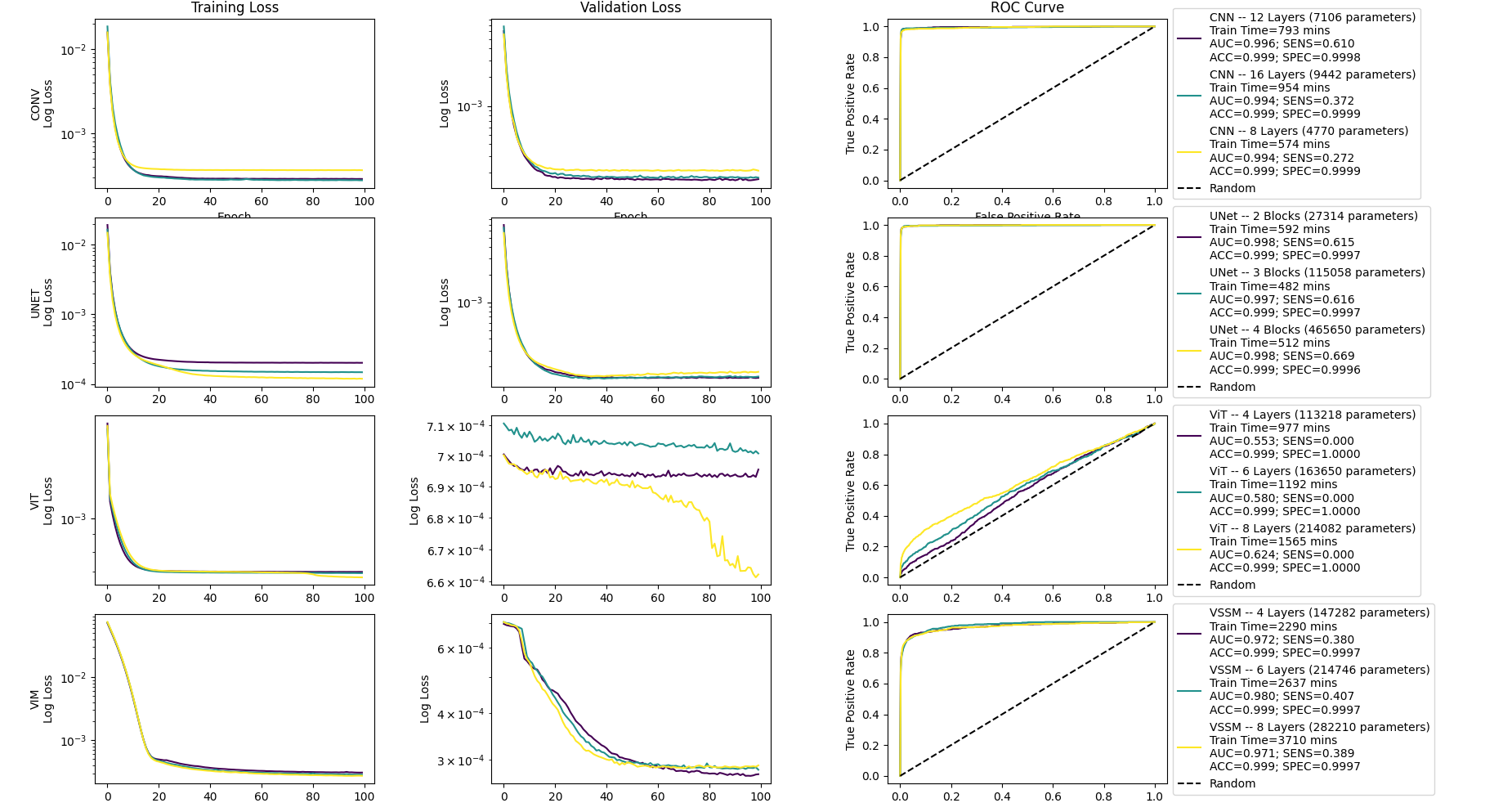}
    \caption{
    \textbf{Comparing model parameters on phase-contrast \textit{Bdellovibrio} data.} Here, each row corresponds to a different model. For each model, we compare results under 3 different parameter choices. The left panel of each row shows training loss per epoch. The middle panel of each row shows validation loss per epoch. The right panel of each row shows the ROC curve. In each panel, we plot the corresponding statistics of each model in a different color. The set of model parameters that corresponds to each color can be found in the legend on the far right, where we also include the hard model performance metrics.}
    \label{SI:fig:bdello_allstats}
\end{figure}

\begin{figure}[H]
    \centering
    \includegraphics[width=\linewidth]{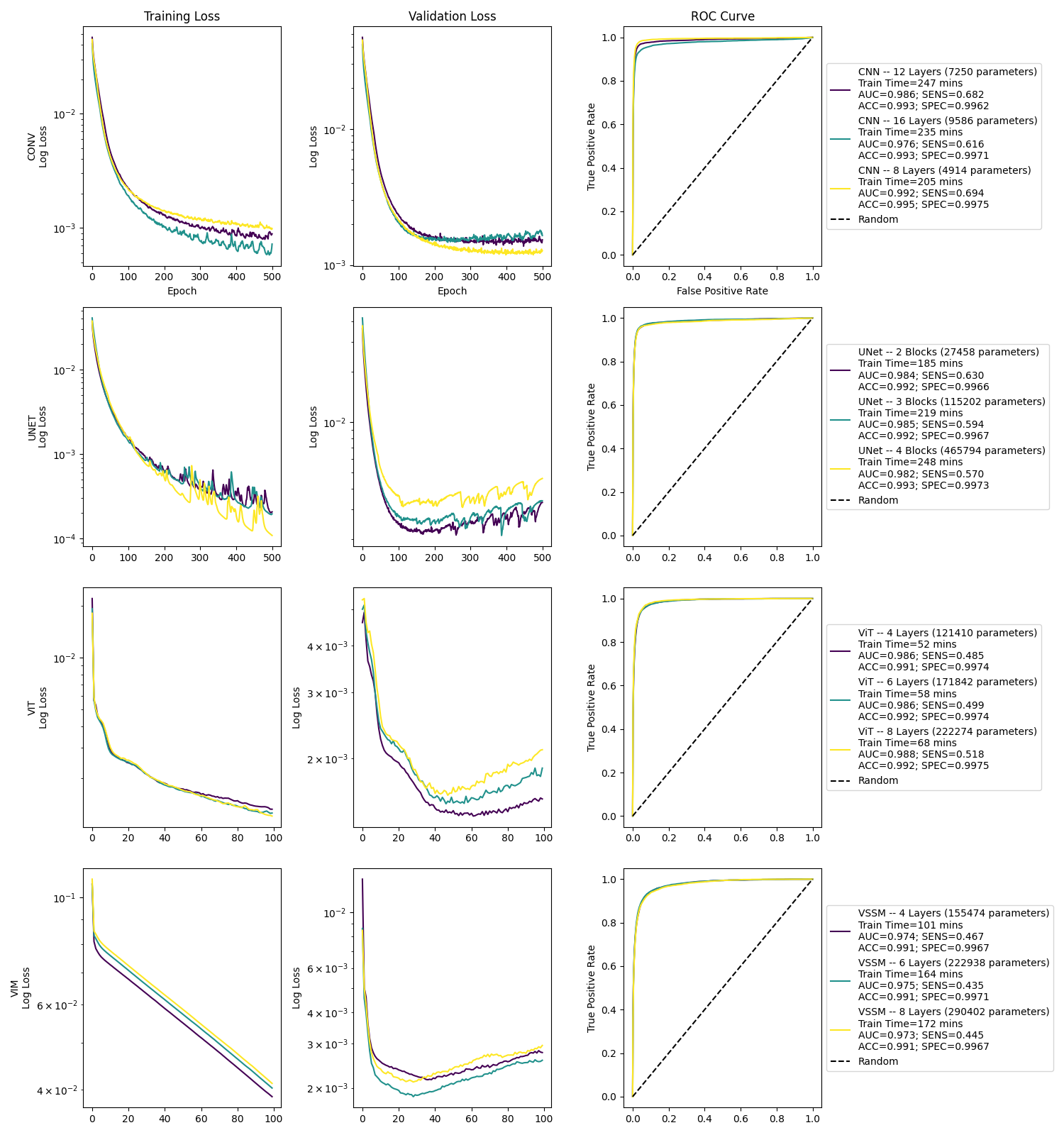}
    \caption{
    \textbf{Comparing model parameters on fluorescence microscopy neuron data.} This figure is the same as figure~\ref{SI:fig:bdello_allstats}, but with results from the fluorescence microscopy neuron dataset.}
    \label{SI:fig:neurons_allstats}
\end{figure}

\begin{figure}[H]
    \centering
    \includegraphics[width=\linewidth]{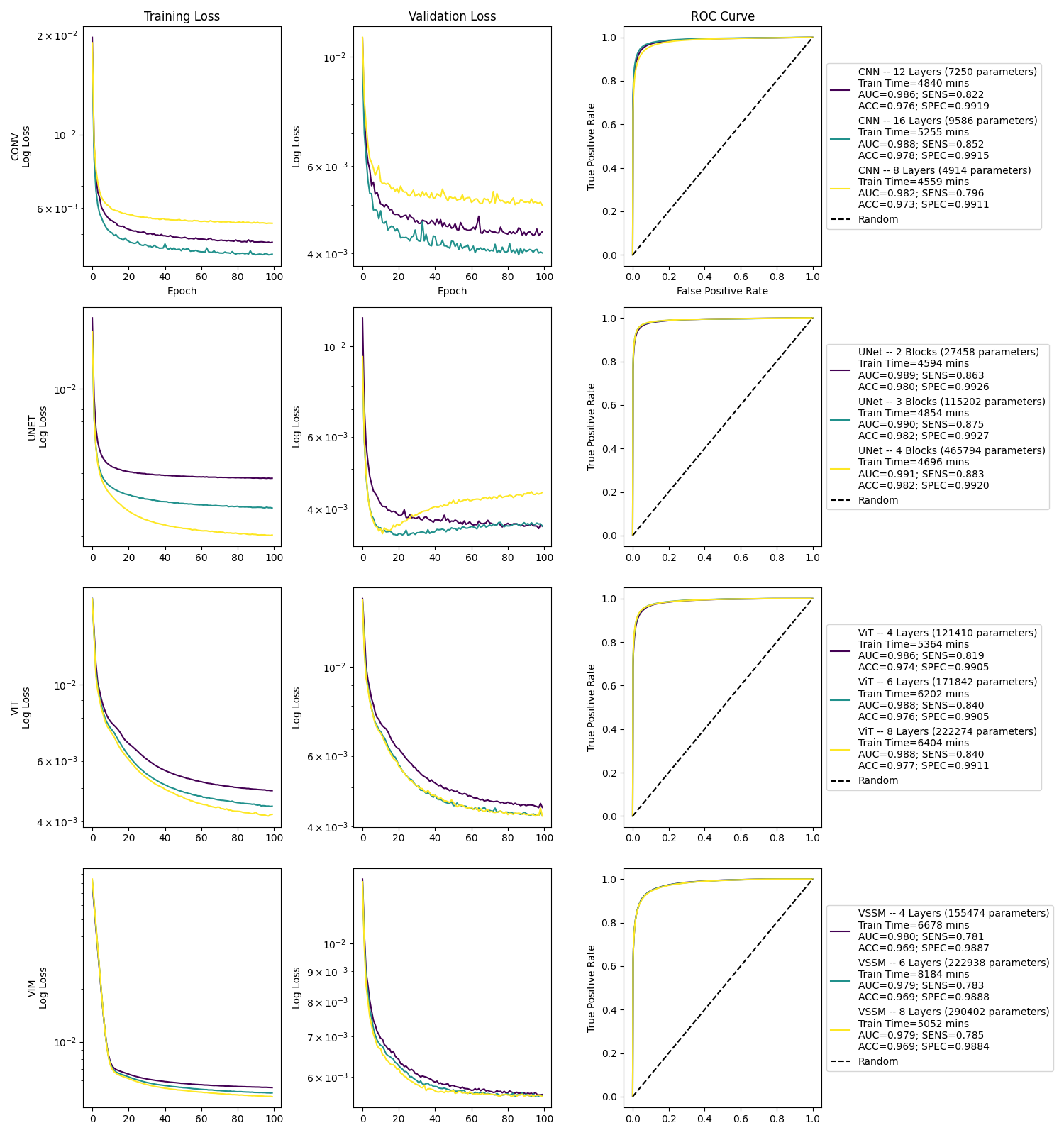}
    \caption{
    \textbf{Comparing model parameters on retina fundus data.} This figure is the same as figure~\ref{SI:fig:bdello_allstats}, but with results on the Retina dataset.}
    \label{SI:fig:retina_allstats}
\end{figure}

Here, we present the hard metrics for model training, comparing each model against itself under different parameter regimes. For the CNN, ViT, and VSSM, we compare different numbers of layers. For the U-Net, we compare different numbers of blocks. The results are shown in Figures~\ref{SI:fig:bdello_allstats}, \ref{SI:fig:neurons_allstats}, and~\ref{SI:fig:retina_allstats}.

In each figure, each row represents a different model type (CNN, U-Net, ViT, VSSM), with each color in the row corresponding to a different parameter value. The left column displays training loss per epoch, the middle column shows validation loss per epoch, and the final column presents the ROC score for each model. The legend on the far right indicates the color coding for each model and lists the values of the hard metrics (AUC score, accuracy, sensitivity, specificity) used in the comparison.

After evaluating the different parameter regimes, the optimal parameters for each model were selected for use in the final comparisons between different models.

% SI Bdello 

% SI Neuron 

% SI Retina 

%% END DOCUMENT
\end{document}